\def\b{\begin{equation}}
\def\e{\end{equation}}
\def\l{\left}
\def\r{\right}
\def\l{\left}
\def\r{\right}
\documentclass[smallextended]{svjour3}     
\smartqed  
\usepackage{graphicx}
\begin{document}

\title{ Einstein energy associated with the Friedmann -Robertson -Walker metric} 

\subtitle{Einstein  energy of FRW metric\\}


\author{Abhas Mitra
}


\institute{A. Mitra \at
              Theoretical Astrophysics Section \\
              Bhabha Atomic Research Centre\\
              Mumbai-400085, India\\
              Tel.: +91-22-25595186\\
              Fax: +91-22-25505151\\
              \email{amitra@barc.gov.in}           
}

\date{Received: date / Accepted: date}

\maketitle

\begin{abstract}
Following Einstein's definition of  Lagrangian density and gravitational field energy density\cite{1,2,3}, Tolman derived a general formula for the total matter plus gravitational field energy ($P_0$) of an arbitrary system\cite{4,5,6}. For a static isolated system, in quasi-Cartesian coordinates, this formula leads to the well known result
$P_0 = \int \sqrt{-g} (T_0^0 - T_1^1 -T_2^2 -T_3^3) ~d^3 x$, where $g$ is the determinant of the metric tensor and $T^a_b$ is the energy momentum tensor of the  {\em matter}. Though in the literature, this is known as ``Tolman Mass'', it must be realized that this is essentially ``Einstein Mass'' because the underlying pseudo-tensor here is due to Einstein. In fact, Landau -Lifshitz obtained the same expression for the ``inertial mass'' of a static isolated system  without
using any pseudo-tensor at all and which points to physical significance and correctness of Einstein Mass\cite{7}!  For the first time we apply this general formula to find an expression for $P_0$ for the Friedmann- Robertson -Walker
(FRW) metric by using the same quasi-Cartesian basis. As we analyze this new result,  physically, a spatially flat model having no cosmological constant is suggested. Eventually, it is seen that  conservation of  $P_0$ is honoured only in the  a static limit.

\end{abstract}

\section{Introduction}
\label{intro}
Einstein was   the first relativist to point out that in curved spacetime matter energy momentum itself is not conserved\cite{1,2,3}. On the other hand, what is conserved is the joint energy momentum of matter and the gravitational field produced by matter itself. This is somewhat like the fact that for a system
of interacting charges (matter), it is the sum of the matter energy momentum and the electromagnetic field energy momentum produced by the charges themselves which is conserved. However, in the presence of  gravitation, because of curved spacetime what is eventually conserved is a sort of geometry weighted
energy momentum density which is now called ``Energy Momentum Complex'' (EMC)\cite{6,8}:
\b
 {\partial \theta_a^b\over \partial x^b} =0
 \e
 The Einstein EMC is given by\cite{3,4,5,6}

\b
 \theta_a^b = \sqrt{-g}~ (T_a^b + t_a^b) 
 \e
 Here $t_a^b$ is the gravitation field energy momentum (pseudo) tensor constructed from  metric tensor $g_{ab}$ and $\partial g_{ab}/\partial x^c$.
 The ``energy complex'' derived from Einstein EMC obviously is
 \b
 \theta_0^0 =\sqrt{-g} ~(T^0_0 + t^0_0)
 \e
 where the subscript $0$ referes to time coordinate.
 The Einstein pseudotensor is often referred to as the ``canonical'' pseudotensor because it is derived by using the general formula for the energy momentum of a classical field with a Lagrangian density and field variables which could be tensors of any rank. In the given case, the field variables are the components of $g_{ab}$. Obviously, because of its the nontensor nature, the local energy density of the field does not have a covariant significance. Indeed,  this formulation of the consevation law got criticized because, in spherical coordinates, empty Minkowski spacetime would appear to possess finite energy density. Einstein tried to answer such criticism in two ways\cite{9}:
  
  (i) He showed that though $t_a^b$ is not a true tensor, the total energy momentum
  \b
  P_a = \int \theta_a^0 ~d^3 x
  \e
    nevertheless behaves as a free vector (i.e., a vector not defined at a particular spacetime point) under linear transforformations for an asymptotically flat spacetime (AFST).
  
  (ii) He also pointed out that, one must evaluate $t_a^b$ and $\theta_a^b$ in quasi-Cartesian coordinates. When this prescription is followed, an empty
  Minkowski spacetime indeed yields nil value for both these quantities (however see later discussion).
  
  Since then, all the authors have recognized the fact that for a meaningful physical answer, the Einstein pseudo tensor must be evaluated in quasi-Cartesian coordinates\cite{6,8}. A case in point would the question of energy momentum flux associated with cylindrical gravitational waves (GW), probably the only known exact solution of GW. For instance, initially Rosen computed the energy momentum flux for this problem using $\theta_a^b$(Einstein) in symmetry adapted cylindrical
  polar coordinates and found nil flux\cite{10}. This led him to conclude that GWs may not be carrying any energy momentum flux and hence, in a sense, pseudo or fictitious quantity. But  following the suggestion  of S. Virbhadra, when Rosen \& Virbharda recomputed the energy flux using quasi-Cartesian coordinates, they found {\em finite value of flux}\cite{11}.
  
  Though Tolman too acknowledged the fact that there is an ambiguity in the localization of gravitational energy momentum, he\cite{4} claimed that
  
  ``It can be shown, nevertheless, that the equations have the necessary fundamental property of being true in all sets of coordinates, and a completely satisfactory justification of the formulation was finally given by Einstein in 1918\cite{9}.''
  
  In the context of the criticism of the energy momentum conservation Eq.(1) which does not involve true tensors, Tolman\cite{4} went on adding that
  
  ``It should be remarked, however, that the appropriate criterion for the fundamental significance of equations should not be that they are written in tensorial form but that they are written in covariant form so as to be true in all sets of coordinates. All tensor equations are indeed covariant equations, but this does not exclude the possibility of covariant equations, such as the above (i.e., Eq.[1]), which are not tensorial. To assume the contrary would be the fallacy of the Dormouse in Alice in Wonderland, who said: ``I breathe when I sleep'' is the same as ``I sleep when I breathe.''
  
  However, the fact remains that even if Eq.(1) may look covariant, Eq.(2) is certainly not so.
  It may be also worthwhile to repeat the well known fact that, since energy momentum conservation after all appears as a divergence, the choice of the pseudo tensor or EMC is by no means unique. In general, the EMC may have a form\cite{8}
  \b
  \theta_a^b (general) = (-g)^{n/2} (T_a^b + t_a^b)
  \e
  where $n$ is a positive integer. An interested reader may look into the forms of some well known pseudo-tensors in the literature\cite{6,8}. This freedom of choosing various EMCs could be something like the gauge freedom in choosing vector potentials in electrodynamics. In particular, the M$\o$ller pseuotensor, by construction, is  a true tensor and thus yields a coordinate independent covariant description\cite{12}. However, this is also the only pseudo-tensor which contains 2nd derivatives of $g_{ab}$.
  
  In general, in view of the non-covariant nature of the pseudo-tensor based energy-momentum localization, the coordinate independent quasilocal approach may be conceptually more important\cite{13}. In this approach, quasilocal energy momentum is obtained from an appropriate Hamiltonian. However, by no means, this approach solves the ambiguity with respect to the assignment of gravitational energy momentum because there could be infinite choices of Hamiltonian boundary conditions.  Further, Chang \& Nester pointed out that the quasilocal approach is intrinsically linked to the pseudotensor approach because each pseudotensor seems to be associated with a legitimate Hamiltonian boundary condition\cite{14}. Thus the importance of the pseuotensor approach
  may not be undermined on the plea that it does not offer a covariant description. As to the criticism, that, all pseudo tensor densities should vanish
  if evaluated in Riemannian normal coordinates where first derivatives of $g_{ab}$ vanish, one may appreciate that such a vanishing could be a necessary consequence of principle of equivalence on which general relativity is based. In particular,  as pointed out by Xulu\cite{6} and as to be pointed out by us later, Einstein pseudo tensor, when evaluated in quasi-Cartesian coordintes,  may indeed yield a  good physical description.

 The field part in the  Einstein EMC is given by\cite{3,4,5,6}
 \b
 \sqrt{-g}~  t_a^b ={1\over 16 \pi} \l( \delta^a_b [L+ 2  \sqrt{-g} \Lambda] - {\partial L\over \partial {g^{pq}_a}} {g^{pq}_b} \r)
 \e
 where   $\Lambda$ is the cosmological constant. Note that Tolman did include $\Lambda$ in this context; for instance see Eq.(87.11) of Ref.\cite{5}.
  Here the Lagrangian density is given by
 \b
 L =\sqrt{-g}~g^{ab} ~\l(\Gamma^p_{ab} \Gamma^q_{pq} - \Gamma^p_{aq} \Gamma^q_{bp}\r)
 \e
 in terms of the Christoffel symbols
 \b
 \Gamma^a_{bc} = {1\over 2} g^{ad} \l( \partial_c g_{db} + \partial_b g_{cd} - \partial_d g_{bc} \r)
 \e
 Note, here, $\partial_a$ denotes differentiation with the coordinate $x^a$. Also,
  the symbol
  \b
  {\partial L \over \partial g^{ab}_c} = - \Gamma_{ab}^c + {1\over 2} \delta^c_a ~\Gamma_{bd}^d + {1\over 2} \delta_b^c ~\Gamma_{ad}^d
  \e
  Recall that the Eqs.(1) and (2) look similar to flat spacetime electromagnetic counterpart\cite{5,6,7}:
 \b
 {\partial \over \partial x^a} \l[T_a^b(m) + t_a^b (f)\r] =0
 \e
 where  $T^a_b (m)$ is due energy momentum tensor of matter  (i.e., charge) and $t^a_b (f)$ is the same for the electromagnetic field.
If we would define
 \b
 P_a = \int \sqrt{-g} ~(T^0_a + t^0_a)~d^3 x
 \e
 where $0$ denotes time coordinate and the integration spans entire 3-space. Tolman\cite{4,5}
  showed that for a finite system with a boundary in AFST
 \b
 {\partial P_a\over \partial t} = Flux ~of ~~ (T^{a\alpha} + t^{a\alpha})
 \e

 Thus,  following Einstein and Tolman, atleast for isolated objects lying in AFST, it seems natural to define $P_a$ as the energy-momentum 4-vector of the system even though $P_a$ is not a true
  covariant vector\cite{1,2,3,4,5,6}. 
  Let us now call
  \b
 P_0 = \int \theta_0^0~d^3 x
 \e
 as the total ``so-called'' Einstein energy of the system. We use this adjective because $P_0$ still depends on coordinates.

  As pointed out by one  anonymous referee, in the presence of a $\Lambda$, one would not have an AFST, and, in a strict sense, the above notion of ``flux'' and associated energy momentum conservation will break down.
 However, by definition  ``universe''  is  the ultimate ``isolated'' object from which {\it there cannot be any outward flux}. Then, we may indeed expect energy momentum of the universe to be conserved by definition. 
 Further since the FRW metric does admit  a  time like (conformal) Killing vector\cite{15,16}, we may expect properly defined total energy to be conserved.  
  
  
 Few years after Tolman's seminal work in 1931\cite{4},  von Freud\cite{17} showed that Einstein EMC could be expressed as the divergence of an appropriate superpotential:
  \b
  \theta_b^a = {1\over 16 \pi} H^{ac}_{b,c}
  \e
  where a comma denotes partial diffentiation and the super-potential is given by

  \b
  H^{a c}_{b} = {1\over \sqrt{-g}} g_{ap} \l[ -g(g^{bp} g^{cq} - g^{cp} g^{bq})\r]_{,q}
  \e 
  One may further note that in terms of this superpotential, Einstein energy complex is given by
  \b
  \theta_0^0 = {1\over 16 \pi} H^{0c}_{0,c}
  \e
  
  In this superpotential formalism, however, the physically important quantity $T_a^b$ does not appear explicitly.  In contrast, as we would see, $\theta_a^b$, {\em when evaluated directly} would contain $T_a^b$ through the Ricci tensor $R_a^b$. In the following we review this formalism and point out that atleast for a static system, the result obtained by this generic formalism matches with one which {\em does not invoke pseudo potential at all}\cite{7}. Finally, we will use this Tolman ansatz to find an expression for $P_0$ for the FRW universe
  directly by retaining the physically important $T_a^b$ term. It may be however be stressed that, the Tolman ansatz does not involve any new psedotensor. On the other hand, it is an ansatz to find $P_0$ {\em directly by using Einstein psedotensor}. Therefore, though, in the literature,
  the resultant $P_0$ is often referred to as the ``Tolman Mass'', we would insist for the term ``Einstein Energy'' for the same.
  
   \section{Einstein - Tolman Formula}
  \label{sec :1}
  Let us recall the Einstein Eq.
   \b
  R_a^b - {1\over 2} R g_a^b + \Lambda g_a^b = - 8 \pi T_a^b
  \e 
  where
 \b 
 R =  8 \pi \l(T_0^0 + T_1^1 + T_2^2 + T_3^3 \r) + 4 \Lambda
 \e
 is the Ricci scalar. 
  By combinging the Einstein Eq. and the expression for $t_a^b$ (Eq.[6]), Tolman showed that (see Eq.89.1)\cite{5}, one can write the Einstein EMC as
  \b
  \theta_a^b (Einstein)  = - {1\over 8 \pi} \sqrt{-g} R_a^b + {1\over 16 \pi} \delta_a^b \sqrt{-g} R - {1\over 16 \pi} g^{pq}_a {\partial L\over \partial g_b^{pq} } + {1\over 16 \pi} \delta_a^b L
  \e
  where $L$ is related to 
   vacuum Einstein Lagrangian density $L_G = \sqrt{-g} R$,  in the following way\cite{4,5,6}:
 \b
 L = \partial_c \l( g^{ab} {\partial L\over \partial {g^{ab}_c}}  \r) - \sqrt{-g} R
 \e
 By inserting this expression for $L$, we finally obtain
 \b
  \theta_a^b (Einstein)  = - {1\over 8 \pi} \sqrt{-g} R_a^b  - {1\over 16 \pi} g^{pq}_a {\partial L\over \partial g_b^{pq} } + {1\over 16 \pi} \delta_a^b \partial_c \l( g^{ab} {\partial L\over \partial {g^{ab}_c}}  \r)
  \e
  It should be noted  that although $\Lambda$ does not explicitly appear in the foregoing relation,
  it remains hidden within the $R_a^b$ term. 
  Thue we obtain
  \b
  \theta_0^0 (Einstein)  = - {1\over 8 \pi} \sqrt{-g} R_0^0  - {1\over 16 \pi} g^{ab}_0 {\partial L\over \partial g_0^{ab} } + {1\over 16 \pi}  \partial_c \l( g^{aa} {\partial L\over \partial {g^{aa}_c}}  \r)
  \e
  Since,
  \b
  R_0^0 = 4\pi (T_1^1 +T_2^2 +T_3^3 -T_0^0) + \Lambda
  \e
 we may split $\theta_0^0$ into 3 terms:

 \b
 \theta_0^0 = A +B +C
 \e
 where
 \b
  A = {1\over 2} \sqrt{-g} ~\l( T_0^0 - T_1^1 - T_2^2 - T_3^3 -{\Lambda \over 4 \pi} \r) =\l({-\sqrt{-g}\over 8 \pi} R_0^0\r),
  \e
  \b
  B={1\over 16 \pi}\l[ \partial_1 \l( g^{ab}{\partial  L \over \partial g_1^{ab}}\r)+\partial_2 \l( g^{ab} {\partial  L \over \partial g_2^{ab}}\r) +\partial_3 \l( g^{ab} {\partial  L \over \partial g_3^{ab}}\r) \r]
  \e
  and
  \b
  C= {1\over 16 \pi} ~ g^{ab} \partial_0 \l( {\partial L\over \partial g_0^{ab}}\r)
  \e

 Since no assumption or precondition has been imposed for the derivation of the above formula, it is valid for arbitrary system, whether
 it has a boundary or not, whether it is static or not and whether it is spherical or not. Note the physical and geometrical significance of
 $A$ in view of the presence of $T_a^b$ and $R_0^0$ in it.
  Thus {\em for any arbitrary system}, spherical or non-spherical, static or non-static, having boundary or no-boundary, one may express
 (see Eq.[23] of \cite{4} and Eq.[92.2] of \cite{5}):
 \b
 P_0 = \int \theta_0^0 ~d^3 x = \int (A + B + C) ~ d^3 x
 \e
 where the integration spans over entire 3-space.

   \subsection{Finite System with a Boundary}
   
\label {sec: 1.1}  
First application of this formula was made by Tolman\cite{4,5} himself for finding the mass-energy of a static finite systen resting in an AFST. Far away from the body, the metric is expected to  assume the form of an exterior Schwarzschild one:
   \b
   e^\nu = e^{-\lambda} = 1 - 2 M/R
   \e
   where $M$ is an integration constant and $R$ is  the circumference coordinate (not to be confused with Ricci scalar).
   Eventually,  at very large $R$, in quasi-Catersian coordinates, one should be able to approximate
   \b
   ds^2 = (1 - 2M/R) dt^2 - (1+ 2M/R) (dx^2 + dy^2 + dz^2)
   \e
   In this AFST, derivation of Kepler's 3rd law shows that, the integration constant $M$ is the gravitational mass of the object provided spacetime is AF. It is clear that the basic requirement for an AFST is not fulfilled if  $\Lambda \neq 0$ because then one would have\cite{5}
     \b
   e^\nu = e^{-\lambda} = 1 - 2 M/R - \Lambda R^2/3
   \e
   Accordingly, in this context, Tolman dropped the $\Lambda$ term (see p.230)\cite{5} 
  and used (quasi) Cartesian coordinates
  with $x^1 =x$, $x^2 =y$ and $x^3=z$. Then he showed that the if the volume integration over $B$ would  be converted into a surface integral at spatial infinity, one would obtain 
  \b
  \int B ~dx dy dz = {1\over 2} P_0
  \e
  so that
  \b
   P_0 = \int A~dx dy dz +  {1\over 2} P_0 +  \int C ~dx dy dz
  \e
  By transposing and then multiplying by $2$, one would obtain
  \b
   P_0 = 2\int A~dx dy dz +   2 \int C ~dx dy dz
  \e
  Further, if the system is static, one has $C=0$. In such a case one would obtain
  \b
   P_0 =  \int \sqrt{-g} ~\l( T_0^0 - T_1^1 - T_2^2 - T_3^3\r)~d^3 x
  \e
  even if the system is not spherically symmetric. 
 Interestingly, for an AFST, Landau \& Lifshitz\cite{7} (pp.348, Eq.[100.19])  {\em obtained exactly the same relation for the total matter plus field of a static system  without invoking any pseudo tensor at all}. This shows the physical correctness of the Einstein EMC
  provided one would evaluate it in quasi-Cartesian coordinates. 
  
  One can see from such exercises by Tolman\cite{4,5} and Landau \& Lifshitz\cite{7}  that for static case, in the absence of $\Lambda$, one may distinctly split $P_0$  into a
   matter contribution
  \b
   P_0(Matter) =  \int \sqrt{-g} ~ T_0^0 ~d^3x,
  \e
  and a pure field contribution 
  \b
  P_0 (field) = \int \sqrt{-g} ~t_0^0 ~d^3x = 
 \int \sqrt{-g} \l(  - T_1^1 - T_2^2 - T_3^3\r)~d^3x
  \e
  																																																																																																																																																																																																																																																																																																																																																																																																																																																																																																																																																																																																																																																																																																																																																																																																																																																																																																																																																																																																																																																																																																																																																																																																																																																																																																																																																																																																																																																																																																																																																																																																																																																																																																																																																																																																																																																																																																																																																																																																																																																																																																																																																																																																																																																																																																																																																																																																																																																																																																																																																																																																																																																																																																																																																																																																																																																																																																																																																																																																																																																																																																																																																																																																																																																																																																																																																																																																																																																																																																																																																																																																																																																																																																																																																																																																																																																																																																																																																																																																																																																																																																																																																																																																																																																																																																																																																																																																																																																																																																																																																											where the integrations now effectively 									span {\em only the region occupied by the matter}. This happens because, in the absence of $\Lambda$, by definition,
  $T_a^b =0$ exterior to the body. Further, if the matter is represented by a perfect fluid, we will have
  \b
T_b^a = (\rho + p) u^a u_b - p \delta^a_b
\e
where $u^a$ is the matter 4-velocity and $p$ is the isotropic (matter) pressure. In the  comoving frame, one obtains 
\b
u^a = \delta_0^a
\e
and the non-vanishing components of $T_b^a$ are ($G=c=1$)
\b
T_0^0 = \rho; \qquad T_1^1 =T_2^2 =T_3^3 = -p
\e
Accordingly, in the comoving frame one has  
\b
   P_0(Matter) =  \int \sqrt{-g} ~ \rho ~d^3x
  \e
and
\b
  P_0 (field) = \int {t_0^0} ~d^3x = \int \sqrt{-g} ~3 p~d^3x
  \e
  
  Much later, Herrera et al.\cite{17,18} used Tolman's formula for finding the {\em active gravitational mass} of a spherically symmetrical collapsing
  fluid in AFST. This was probably the maiden application of Tolman's generic prescription for a non-static system having a boundary. On the other hand,
  we will invoke this formalism to find the energy of a non-static system {\em not having any boundary at all}.
  	
  \subsection{Inclusion of $\Lambda$ For Isolated Objects?}
  If the $\Lambda$ term would be shifted to the right hand side of Einstein Eq., it follows that, in the comoving frame, one may replace
  $\rho$ and $p$ by their respective {\em effective values}:
  \b
\rho_e = \rho + \Lambda/8 \pi; \qquad p_e = p - \Lambda/8 \pi
\e
  Then one might think that, in such a case,  the definition of the Einstein mass for a static isolated system would be modified into
  \b
P_0 = \int \sqrt{-g} (\rho_e + 3 p_e) ~d^3x
\e
  But it may not be so because of the following reasons. As seen by the exterior metric, when $\Lambda$ is present, in a strict sense, the spacetime ceases to be
  asymptotically flat.  Since,  now, there would be no {\em strict} Kepler's 3rd law, one cannot identify the integration constant $M$ with Schwarzschild mass. However it might still be possible  to define ``mass'' by using more advanced and specific ideas which is beyond the scope of this discussion\cite{13}.
  
  Nevertheless, it may be noted that the spacetime  appears to acquire, strange and rather unphysical properties. Because it would be seen that 
  while $g_{00}$ initially monotonically increases with $R$, it would start decreasing subsequently. And at $R=R_*$, defined by the roots of the following Eq.
  \b
  g_{00} =1- {2M\over R{_*}}  - \Lambda {R{_*}^2\over 3} =0,
  \e
  one would have a {\em singular} situation with $g_{00} =0$! And following this, $g_{00}$ would reverse its sign to approach $-\infty$!! Since the gravitational red/blue shift of the photons emitted by the body depends on the value of the $g_{00} (R)$ at the point of observation, one would see
  strange and abnormal red/blue shifts.
   At this juncture, one would argue that, for sufficiently, large $R$, the metric of the {\em isolated} object must merge with cosmological metric, which under, the assumption of homogeneous and isotropy does not admit any gravitational red/blue shift. But the fact remains that, any mass determination might get  swamped by a cosmological vacuum contribution
  \b
  P_0 (vacuum) \sim \int \sqrt{-g} ~(- \Lambda/4\pi) ~d^3x
  \e
  which would diverge if the spatial section would be of infinite extent. Irrespective of the precise boundary condition, the foregoing Eq. indicates that, a completely vacuum universe should posses {\em a net negative energy density} because of the negative contribution due to $p_e = - \Lambda/8 \pi$!
  We would seek a more precise answer to this question later.

  \section{Negative Self-gravitational Energy}
  If the metric determinant of the spatial section is $-h$, one can write
  \b
  g = g_{00} (-h)
  \e
  and accordingly, following Landau-Lifshitz derivation of $M$ for an isolated static system, we may write
  \b
P_0 = M = \int \sqrt{g_{00}} (\rho + 3 p ) ~d{\cal V}
\e
  where
  \b
  d{\cal V} = \sqrt{-h} ~d^3 x
  \e
  is proper 3-volume element. In modern terminology, $P_0 = M$ is the ADM mass measured by a faraway inertial observer $S^\infty$. Thus  the comoving (local) Active Gravitational Mass Density (AGMD)
  \b
  \rho_g = \rho + 3p
  \e
  indeed appears to increase due to the ``weight'' of pressure,
  It is  however important to note that {\em this pressure contribution is actually  due to the field energy contribution} (when computed in quasi-Cartesian coordinates): $3p = t_0^0$ and the  field energy density   is positive as long as $p$ is positive. This is similar to the fact that while electrostatic {\em interaction} energy can be positive or negative depending on the sign of the electric
charges, the {\em field energy density $E^2/8\pi$} is always positive\cite{7}. Since gravitational waves carry positive energy\cite{20,21,22,23}, in
general, we expect $t_0^0$ to be positive.

So the question which might arise is that then how one obtains {\em negative} self-gravitational energy? Note that the distant inertial observer $S^\infty$, however,  sees an effective total mass density
  \b
  \rho_g^\infty = \sqrt{g_{00}} (\rho + 3p) < \rho_g
  \e
  because $g_{00} <1$. And it is the difference between $\rho_g$ and $\rho_g^\infty$ which gives rise to negative self-gravitational energy when viewed globally\cite{24}.
  This question may be probed by noting that in the zeroth order post  Newtonian limit, the Newtonian gravitational potential is obtained from
  \b
  \psi = {1\over 2} (g_{00} -1) \approx {\nu\over 2}
  \e
  Since the potential $\psi$ is negative, its coupling to the rest of the matter {\em gives rise to global negative self-gravitational energy}.
  Further, it has been explicitly been shown by Tolman (pp. 248-250)\cite{5}, that
  \b
  P_0 = M \approx  \int \rho ~d{\cal V} + \int {1\over 2} \rho \psi ~d{\cal V} =\int \rho ~d{\cal V} - 3 \int p~d{\cal V} =\int(\rho- 3p) d{\cal V}
  \e
  Thus, in this  limit
  \b
  E_g \approx  - \int 3 p d{\cal V}
  \e
  and due to global gravitational coupling, the effective AGMD seen by $S^\infty$:
  \b
  \rho_g^\infty \approx \rho - 3p < \rho
  \e
Hence  one has negative self-gravitational energy only when  $g_{00} <1$ and in general
  $g_{00} =g_{00} (R)$.  In such a case,   as light propagates through a medium having negative self-gravitational energy, one would see gravitational redshift. But, in case, one would fix $g_{00} =1$, i.e., synchronize all clocks in a rather Newtonian fashion, one would not have
  any global self-gravitational energy. In such a case, neither would one see any gravitational red-shift quite like the Newtonian case. All red-shifts
  then must be due to kinematical/Doppler origin. Cosmological redshifts, in the paradigm of the FRW model, are indeed believed to be of purely
  kinematical origin because there is no gravitational redshift for $g_{00} =1 = fixed$. 
  
  For later requirement, let us ponder,  is it possible that $P_0 \to 0$ ever? As hinted by Eq.(48), in principle, it is possible in the singular 
  limit $g_{00} \to 0$.  Only in such a case, in view of Eqs.(53) and (55), one may conceive that positive energy of matter/radiation/heat gets nullified by the negative self-gravitational energy even though field energy density $t_0^0$ remains positive. Note, the field energy $t_0^0$  defined locally via Eq.(6) must not be confused with self-gravitational
  energy $E_g$. For the latter there is no proper local definition. Thus even if such a vanishing of $P_0$ would take place, $\rho$ and $p$ would still retain their original sign
  and there cannot be any cancellation of matter energy with field energy.  In contrast,  (negative) self-gravitational energy is essentially a global concept
   and cannot be defined at a spacetime point. Further, it is mediated by $g_{00} =g_{00} (R) <1$. And when $g_{00} =1$, one should have $E_g =0$ and then there would be no question of cancellation of matter energy. Thus, in the ansence of a variable $g_{00} <1$, an occurrence of $P_0 =0$ should
   signify $\rho =0$.

 \section{ FRW Metric in Cartesian Coordinates}
  \label {Sec :2}
 The isotropic form of the FRW metric is\cite{5}(pp. 338) \cite{8}:
\b
ds^2 = dt^2 - {S^2(t)\over {(1+ k r^2/4)^2} } [d r^2 +  r^2 (d\theta^2 + \sin^2 \theta d\phi^2)]
\e
where $k$ can assume values of $0$, $+1$, or $-1$. 
 This  form can easily be written in terms of Cartesian coordinates:
\b
ds^2 = dt^2 - {S^2(t)\over (1+ k r^2/4)^2}  (dx^2 + dy^2 + dz^2)
\e
where 
\b
r^2 = x^2 + y^2 + z^2
\e
Thus,
\b
g_{11} = g_{22} = g_{33} = - {S^2 \over f^2}; \qquad g_{00} =1
\e
and
\b
g^{11} = g^{22} = g^{33} = - {f^2 \over S^2}; \qquad g^{00} =1
\e
where
\b
f(r) = 1+ k r^2/4
\e
At the beginning,  let us mention that greek letters would represent $x,y,z$ and in this section, repetetion of a greek letter in any expression {\em will not imply any summation}. Then it may be noted that
\b
\partial_\alpha f =   {1\over 2} k x^\alpha
\e

while
\b
\partial_0 f =0
\e
Because of the symmetry we also have
\b
\partial_\alpha  g_{\beta \beta}= p_\alpha=  {k x^\alpha S^2\over f^3} 
\e
\subsection{Useful Christoffel Symbols}
In this quasi Cartesian basis, one may verify that since $g_{00} =1$, the non-vanishing connection coefficients are

\b
\Gamma^\alpha_{0\alpha}  = {1\over 2}g^{11}  q = { {\dot S} \over S}
\e
where
\b
q  = \partial_0 g_{11} = {- 2 S {\dot S} \over f^2}
\e
Next,
\b
\Gamma^0_{\alpha,\alpha} = {-1\over 2} q = {S {\dot S} \over f^2}
\e\footnote{Note, in the published version, there were typos in Eqs.(65) and (67)}.
Further,
\b
\Gamma^\alpha_{\alpha \alpha} =   -{1\over 2} {kx^\alpha\over f}
\e

Also,
\b
\Gamma^2_{12} =  \Gamma^3_{13} =  -{1\over 2} g^{11} p_1 ={1\over 2} {kx\over f}
\e

All other connection coefficients vanish.

 \section{Direct Computation of $P_0$ for FRW Universe}
 \label {Sec: 3}
 Let us first compute the $B$-term:
 \subsection{Evaluation of $B$}
 From Eq.(9), it is seen that 
 \b
 {\partial L \over \partial g_1^{11}} = - \Gamma^1_{11} + {1\over 2} \Gamma^d_{1d} + {1\over 2} \Gamma^d_{1d}= - \Gamma^1_{11} +  \Gamma_d^{1d}  
  =\Gamma^2_{12} + \Gamma^3_{13} = {kx\over f}
 \e
  On the other hand, we have 
 \b
 {\partial L \over \partial g_1^{22}} = - \Gamma^1_{22} = {\partial L \over \partial g_1^{33}} = - \Gamma^1_{33} = {\partial L \over \partial g_\alpha^{00}} = 0
 \e
 
 Then one finds that
 \b
  g^{ab} {\partial L \over \partial g_1^{ab}} =  g^{11} \l({\partial L \over \partial g_1^{11}} + {\partial L \over \partial g_1^{22}} + {\partial L \over \partial g_1^{33}} \r) + 
  {\partial L \over \partial g_1^{00}} = g^{11} {kx\over f} = {-k x f\over S^2}
 \e
 In general, we can write
 
 \b
 g^{ab} {\partial L \over \partial g_\alpha^{ab}} =  {-k x^\alpha f\over S^2}
 \e
By differentiating the foregoing Eq., we obtain,
 \b
 \partial_\alpha \l(g^{ab} {\partial L \over \partial g_\alpha^{ab}}\r) = {-k \over  S^2} \l( f+ {1\over 2} k {x^\alpha}^2 \r)
 \e
 
 By adding the three components of the foregoing equations and also by recalling Eq.(26), we obtain
 \b
 B = {-k \over 16 \pi S^2} \l( 3f+ {1\over 2} k r^2 \r)
 \e
 Now plugging in  the value of $f$ from Eq.(61) in this equation, we finally find
 \b
 B = {-k \over 16 \pi S^2} \l( 3 + {5\over 4} k r^2 \r) 
 \e
 Further since the spatial curvature of the spacetime is
 \b
 K = {k\over S^2}
 \e
 
 we may also write $B$ as
 \b
 B = {-K \over 16 \pi } \l( 3 + {5\over 4} k r^2 \r) 
 \e

 \subsection{ Evaluation of $C$}
 \label {sec: 3.1} 
 From Eq.(9), one can see that when $a=b$, one has
 \b
 {\partial L \over \partial g_0^{ab}} = - \Gamma^0_{aa} + \Gamma^d_{0d}\qquad (no ~sum ~over~ 'a')
 \e
 from which one obtains
 \b
 {\partial L \over \partial g_0^{00}} = \Gamma^1_{01} + \Gamma^2_{02} + \Gamma^3_{03} = {3\over 2} g^{11} q = 3{{\dot S} \over S}
 \e
 and
 \b
 {\partial L \over \partial g_0^{11}} =  -\Gamma^0_{11}  = +{1\over 2} q = {- S {\dot S}\over f^2}
 \e
 Similarly
 \b
 {\partial L \over \partial g_0^{22}} = {\partial L \over \partial g_0^{33}} = {- S {\dot S}\over f^2}
 \e
 Further note that
 \b
 \partial_0 \l({\partial L \over \partial g_0^{00}}\r) =3\l({{\ddot S}\over S} - {{\dot S}^2\over S^2}\r)
 \e
 and
 \b
 g^{11} \partial_0 \l({\partial L \over \partial g_0^{11}}\r) ={ S {\ddot S} + {\dot S}^2\over S^2} = {{\ddot S}\over S} + {{\dot S}^2\over S^2}
 \e
 Similarly,
 \b
 g^{22} \partial_0 \l({\partial L \over \partial g_0^{22}}\r)= g^{33} \partial_0 \l({\partial L \over \partial g_0^{33}}\r)  = {{\ddot S}\over S} + {{\dot S}^2\over S^2}
 \e
 Since $g^{11} = g^{22} = g^{33}$, from Eq.(27), we can write
 \b
 C = {1\over 16 \pi} \l[\partial_0 \l({\partial L \over \partial g_0^{00}}\r) + 3 g^{11} \partial_0 \l({\partial L \over \partial g_0^{11}}\r)\r]
 \e
 By using Eqs.(83-85) in the foregoing Eq., we find that eventually $C$ gets simplified as
 \b
 C = {3\over 8 \pi} {{\ddot S} \over S} 
 \e
 Now let us recall  the evolution Eq. of the FRW universe
 
 \b
 {{\ddot S} \over S} = -{4 \pi\over 3} \rho_*
 \e
 where
 \b
 \rho_* =  (\rho_e + 3p_e) =  (\rho + 3p) - {\Lambda \over 4}
 \e
 Eqn. (88) shows that, even for the FRW case for which neither AFST nor any other boundary condition has been assumed, $\rho_*$ appears to be the AGMD.
 Then by using Eqs.(23) (87) and (88), we can reexpress
 \b
 C = {-\rho_*\over 2} = {R_0^0\over 8 \pi}
 \e
 Clearly, $C$ has a distinct {\em physical and geometrical} meaning in terms of  $\rho_*$ and $R_0^0$.
 \section{ Final form of Einstein Energy Complex}
 \label {sec : 3.2}
 Since for the metric (57), 
 \b
 \sqrt{-g} = {S^3 \over f^3}
 \e
 and the FRW metric is actually a comoving metric, from Eqs. (23) and (25), we obtain
 \b
 A = {S^3 \over  2 f^3} (\rho_e + 3p_e) = {1\over 2} \rho_* {S^3 \over   f^3} =\l({-R_0^0\over 8 \pi}\r) \l({S^3 \over f^3}\r) 
 \e
 By using Eq.(88), we may also write
  \b
 A = {3\over 8 \pi} \l({{-\ddot S} \over S}\r) \l({S^3 \over   f^3}\r)
 \e
Clarly, one would have $A=0$ for a static case.     It is seen that like the static case, for this dynamic case too $A$ has a clean
physical and geometrical meaning in terms of $\rho_*$ and $R_0^0$. The additional factor $S^3/f^3$ is the ratio of  the proper volume element to the coordinate volume element and which becomes unity for $k=0$ in the static case. In view of this assumed dynamic metric, eventually, $A$ behaves quite unlike the case of an isolated object in an aymptotically flat space time
where in the static case $\int A ~d^3 x= P_0/2$\cite{4,5}. 
After combining all the contributions,  the final form of Einstein energy complex $P_0$ for the FRW metric, when evaluated directly,   becomes
 \b
 \theta_0^0 =  {-k \over 16 \pi S^2}  \l( 3 + {5\over 4} k r^2 \r) +
  {3\over 8 \pi} {{\ddot S} \over S}  \l(1- {S^3 \over   f^3} \r) ={-k \over 16 \pi S^2}  \l( 3 + {5\over 4} k r^2 \r) -{\rho_*\over 2} + {\rho_*\over 2} {S^3 \over f^3}
  \e
  In particular, note that, when $k=0$ and $f=1$, one has 
  \b
 \theta_0^0 =  
  {3\over 8 \pi} {{\ddot S} \over S}  (1- S^3 ) = {-1\over 2} \rho_* (1-S^3)
  \e
 Due to spherical symmetry, we may take the coordinate volume element to be $dV = 4\pi r^2 dr$ so that the form of Einstein energy of the FRW metric becomes
  \b
  P_0 = \int 4\pi r^2  \theta_0^0  ~dr
 \e
 By inspecting, the metric (56), it becomes clear that the range of $r = 0, \infty$ for $k =0,1$ while $r=0, 2$ for $k= -1$.
 Further, the proper  volume element is
 \b
 d{\cal V} = {S^3\over f^3} ~dV
 \e
 Now using Eqs.(94-97) we may simplify $P_0$ as
 \b
 P_0=  {-k \over 16 \pi S^2} \int  \l( 3 + {5\over 4} k r^2 \r) dV  + {\rho_*\over 2} ({\cal V} -V)
  \e
 Here the term $\rho_* {\cal V}$ looks like some sort of  effective proper energy content. The 1st term of the right hand side of the foregoing Eq.
 shows the contribution due to {\em curvature of the spatial section alone} while the 2nd term denotes contribution due to {\em spacetime curvature} in general. In particular, the 2nd term implies a contribution with respect to a flat spacetime having no energy. Thus Einstein energy has a nice physical and geometrical interpretation even in this dynamic case. Such a {\em physical and geometrical interpretation of the energy of the FRW metric  has not been revealed before}. In a trivial case of a flat spacetime with $k=0$ and ${\cal V} =V$, obviously $P_0 =0$.

 \section{Analysis of this general formula}
 
 \subsection {Position Dependence of Net Energy Density}
  The first apparently anomalous thing about the Einstein energy complex (94) is that, {\em it is position dependent unless} $k=0$. One  may try to explain away this position dependence by telling that ``gravitational field energy is not localizable due to principle of equivalence''. But all that this latter statement means is that at a given spacetime point, the value of $t_0^0$  would vary as one would employ different coordinate systems and in Riemann normal coordinate system, one would have $t_0^0 =0$. It is in this sense that, one may not demand conservation of energy momentum in a covariant way. But once we have  decided to work  in a given coordinate system, which in this case is the quasi-Caresian coordinate system, and not to compare the values of $t_0^0$ obtained in various coordinate systems, we have already compromised for such non-covariance. Such non-covariance however does not at all imply that a quantity which is expected to be spatially uniform can turn non-uniform. It is true that one speaks of non-localization of position of an electron/photon in quantum mechanics because of inherent wave-particle duality. But we are not treading into any quantum mechanics here. Hence  we do not expect
 $\theta_0^0$ to have any inherent spatial spread . Consequently,  in a {\em supposed isotropic and homogeneous universe}, one should not expect any position dependence of $\theta_0^0$ irrespective of its the precise physical significance.
  
 Further, the Einstein pseudo-tensor need not be be blamed for such an occurrence which seems to defy the assumption of homogeneity and isotropy.
 The real reason for such unexpected position dependence is the position dependence of the metric coefficients $g_{\alpha \alpha}$. This position dependence vanishes only for $k=0$ when $\theta_0^0$ too  becomes position independent. The situation here could be something like the following:
 If one would consider a spherically symmetric finite fluid of constant $\rho$, the pressure still would show position dependence. This position dependence of $p$ can be eliminated by either (i) pushing the boundary to $\infty$, or (ii) by setting $p=0$ by hand, or (iii) by pursuing the spatially flat Newtonian limit $\rho \to 0$. This would be clear as we would discuss below the case of a supposed static universe.

 \subsection {Static Universe}
 In the static case, the expression for Einstein energy complex becomes
 \b
 \theta_0^0 =  {-k \over 16 \pi S^2}  \l( 3 + {5\over 4} k r^2 \r) 
  \e
  It is known that, by means of the following coordinate transformation
  \b
  {\bar r} = { r \over 1 +  k r^2/4}
  \e
  the metric (56) would transform as
  \b
  ds^2 = dt^2 - S^2 \l[ {d{\bar r}^2 \over 1 - k {\bar r}^2} + \bar{r}^2 (d\theta^2 + \sin^2 \theta d\phi^2)\r]
  \e
  For simplification, we now drop bar from ${\bar r}$ to rewrite the above Eq. as 
  \b
  ds^2 = dt^2 - S^2 \l[{ d  r^2 \over 1 - k  r^2} + r^2 (d\theta^2 + \sin^2 \theta d\phi^2)\r]
  \e
  If we recall that the {\em static} interior Schwarzschild solution for a constant density may be expressed as\cite{5,25}
  \b
  ds^2 = e^\nu dt^2 - S^2 \l[{ d  r^2 \over 1 - k  r^2} + r^2 (d\theta^2 + \sin^2 \theta d\phi^2)\r]
  \e
  where $k=+1$ corresponds to an effective $\rho_e >0$, $k=0$ corresponds to $\rho_e =0$ and $k=-1$ corresponds to $\rho_e <0$,
  it might appear that, the FRW metric is similar to a special interior Schwarzschild solution 
  where, somehow, one has
  \b
  e^\nu =1
  \e
  In fact, for exploring a static FRW universe, Tolman indeed arrived at Eq.(102)  by starting from Schwarzschild interior solution (103) (see pp. 333-337)\cite{5}.
  He found that the condition for having $e^\nu =1$ is either
  \b
  \rho_e + 3p_e =0
  \e
  or
  \b
  \rho_e + p_e =0
  \e
  Essentially, Tolman probed the condition  for $e^\nu$ to become position independent.
  On the other hand,  we have probed the question from a broader perspective where all physically meaningful quantities such as the 
  the invariant/scalar acceleration
   experienced by an interior fluid element\cite{25} 
   \b
   a= \sqrt{-a_i a^i} 
   \e
   where 
   \b
   a^i = u^j \nabla_j u^i
   \e
   is the 4-acceleration, 
   and radial pressure gradient $p'$ should {\em indeed be position independent}. We found the following relations which, in general, suggest position
   dependence of $a$ and $p'$:
   \b
   {a\over \rho_e + 3 p_e} = {k S r\over \sqrt{1- kr^2}}
   \e
   and
    \b
   {p'\over (\rho_e +p_e) (\rho_e + 3p_e)} = {k S r\over 1- kr^2}
   \e
   In a homogeneous and isotropic universe, the left hand side of the two foregoing equations must be independent of $r$. But the inspection of the right hand side of the same equations tells that, it could be so only when $k=0$. For a static universe, occurrence of $k=0$ corresponds to $\rho_e =0$. It was independently shown that  in case one would assume $k=1$,  the condition that $a$ and $p'$ would not blow up at $r=1$ demands that both Eqs.(105) and (106) are satisfied\cite{25}.
 This again leads to $\rho_e  =0$. Under weak energy condition, this would mean that a static unverse should have $\Lambda =0$ as well as
   {\em mean} matter density $\rho =0$. Incidentally, occurrence of a mean $\rho =0$ need not imply absence of matter in cosmological models. 
   For example, one has mean zero matter density in an infinitely hierarchial fractal universe\cite{26}. But for isotropy and homogeneous, case, a mean $\rho =0$ may occur only when matter patches of the universe are separated by infinite distances from one another\cite{27,28}. In any case 
   it appears that in order that $\theta_0^0$ is  position independent as per the assumption of isotropy and homogeneity, only the $k=0$ FRW model is physically admissible. Note this argument is irrespective of the question whether $P_0$ is conserved or not.

  \section{Empty Universe with a Positive $\Lambda$}
  As is known, for an empty universe  FRW solution becomes de-Sitter solution 
  \b
  ds^2 = dt^2 -S^2 (dx^2 +dy^2 +dz^2)
  \e
  having $k=0$ and
  \b 
  S = e^{Ht}
  \e
  where
  \b
  H =\sqrt {\Lambda/3}
  \e
  Since $\rho = p= 0$, in this case,  Einstein energy complex becomes
  \b
  \theta_0^0 = {3\over 8 \pi} {\ddot S\over S} (1 -S^3)
  \e
  Using the fact that now
  \b
  {\ddot S\over S} = {\Lambda \over 3}
  \e
  we obtain
  \b
  \theta_0^0 = {\Lambda \over 8 \pi} (1 -S^3)
  \e
  But expcept for the instant of $t=0$, one would have $S >1$ and accordingly, $\theta_0^0 < 0$. Then, one finds
  \b
  P_0 = \int 4 \pi r^2 \theta_0^0  dr= {\Lambda \over 2 \pi} (1 - S^3) \int_0^\infty r^2 dr = {\Lambda} (-\infty)
  \e
  Again occurrence of a negative $\theta_0^0$ and $P_0$ may be related to the fact Active Gravitational Mass Density of pure vacuum $\rho_e + 3p_3 $ is always negative for a positive $\Lambda$. However, from the concept of positivity of mass-energy, occurrence of a negative $P_0$ (and that too $-\infty$) is unphysical, and this would again demand that $\Lambda =0$!
  
  Further, even if one would accept the apparent result that $P_0 = -\infty$, it may be noted that there is latent temporal dependence of $P_0$ if
  $\Lambda \neq 0$:
  \b
  {dP_0\over dt} = {d\theta_0^0 \over dt} \int_0^\infty 4 \pi r^2 dr = - {\Lambda \over 4 \pi} S^3 H (+\infty)
  \e
  Such an occurrence would again suggest that $\Lambda =0$ if we would expect $P_0$ to be conserved.
 
 There could be yet another reason for which one would expect $\Lambda =0$ in the context of de-Sitter solutions. With suitable coordinate transformations, the metric, can be transformed into a {\em static} form (see pp. 346)\cite{5}:
  \b
  ds^2 = (1- R^2/S^2) dT^2 - (1- R^2/S^2)^{-1} dR^2 - R^2 (d\theta^2 + \sin^2\theta d\phi^2)
  \e
  In fact, de-Sitter originally obtained his solution in the foregoing form. As such, there is nothing anomalous in the fact that a non-static metric
  may look static after a coordinate transformation which involves time. But such form changes should not induce physical changes. Note in the foregoing metric, the radial coordinate $R$ is the circumference coordinate which is a scalar and directly related to luminosity distances. Since now
  $g_{00} = g_{00}(R)$, one would expect that moving photons would {\em experience gravitational redshift}. But as per the FRW metric, there should not be any gravitational red-shift! Such a physical contradiction can be removed by realizing that $\Lambda =0$ and de-Sitter metric is actually flat Minkowski metric.
  
  \section{Time Dependence of Einstein Energy}
  We have already discussed the issue of time-dependence of $\theta_0^0$ and $P_0$ for $k=0$ model in the context of de Sitter case. For more specific discussion, let us first note that for $k=1$, $f = 1 + r^2/4$ we will have
  \b
  {\cal V} = 4 \pi\int_0^\infty  {S^3 \over f^3} r^2 dr  = 4 \pi S^3 \int_0^\infty   {r^2 \over (1 + r^2/4)^3} dr = 2 \pi^2 S^3,
  \e
  where ${\cal V}$ is proper 3-volume. On the other hand, the coordinate volume is
  \b
  V =\infty
  \e
 Also,
  \b
  I_1= 4 \pi \int_0^\infty  r^2 [3 + (5/4)r^2] dr = \infty
  \e
  On the other hand, for $k =-1$, $f= 1 - r^2/4$ and
  \b
  {\cal V} = 4 \pi\int_0^2  {S^3 \over f^3}r^2 dr = 4 \pi S^3 \int_0^2  {r^2 \over (1 - r^2/4)^3} dr = S^3 \infty=\infty,
  \e
  \b
  V = \int_0^2 4 \pi r^2 dr = {32 \pi\over 3}
  \e
  and
  \b
  I_2= \int_0^2 4\pi r^2 [3 - (5/4)r^2] dr = 0
  \e
  Then  we find that
  \b
  P_0 = 0 + {3\over 8 \pi} {\ddot S \over S} (1-S^3) (\infty) = -(\rho_*/2) (1- S^3)(\infty); \qquad k =0,
  \e
  
   \b
  P_0 = 0 + {3\over 8 \pi} {\ddot S \over S} (32\pi/3 -S^3) (\infty)  = (-\rho_*/2) (32 \pi/3 -S^3) \infty ; \qquad k = -1
  \e
  and
  
  \b
  P_0 = -{\infty\over S^2} + {3\over 8 \pi} {\ddot S \over S} ( \infty - 2 \pi^2 S^3 )  =-{\infty\over S^2} -(\rho_*/2) (\infty - 2\pi^2 S^3); \qquad k = +1
  \e
  In case, one would consider, $S = finite$ for the $k=1$ case, one would obtain
  \b
  P_0 = -\infty - \rho_* (\infty); \qquad k= +1; ~~ S=finite
  \e
  In this case, $P_0$ could be $+\infty$ or $-\infty$ depending on the sign and value of $\rho_*$.
  
  On the other hand, if $S=\infty$ for the above case, one may have
  \b
  P_0 = 0
  \e
  Since, from symmetry conditions, it has been argued that one should have, $P_0 =0$ for a closed universe\cite{7,29}, we find that, the suposed closed case
  should actually be an open case with $S =\infty$.
  In general,  $P_0$  would be time dependent if $\rho_*$ would be so. For $k =-1$ case too, $P_0$ could be either $\infty$ or $-\infty$ unless $\rho_* =0$. Same is true for the simplest $k=0$ case. We have found that for an isolated static object in an AFST, $P_0$ yields a result (in quasi-Cartesian coordinates) which is actually obtainable without invoking psedotensor at all. Thus, if we would consider that $P_0$ is a physically meaningful quantity
  irrespective of the question whether it is conserved or not and it cannot fluctuate wildly between $+\infty$ and $-\infty$ (even for a supposed closed geometry), we should adopt
  \b
  {\ddot S}/S =0; \qquad \rho_* =0;  ~~S=\infty
  \e
  There may be a simple mathematical reason as to why one should have $S=fixed$. It is known that, if indeed $S=S(t)$, the spacetime may be extrapolated back to a {\em geometrical point}. But what could be the proper 3-volume of a point: ${\cal V} (point) = ?$. Since this is expected to be
  ${\cal V} =0$, a time dependent open FRW model is immediately ruled out because in the latter case, ${\cal V} =\infty$!
  And of course, if we would demand that $P_0$ is conseved, the foregoing constaints would be naturally imposed on the FRW model.

  \section{Discussions}
  For the first time, we obtained an expression for the  Einstein energy complex  $\theta_0^0$ for the FRW metric in a {\em direct fashion} i.e.,  without resorting to any superpotential. This direct treatment which explicitly involved $T_a^b$ yielded a direct interpretation of $P_0$ in terms of spacetime curvature and AGMD. The non-covaiant nature of $\theta_0^0$  was handled by working it out in the quasi-Cartesian coordinates
  and by making no demand that its value would be same in other coordinate systems. Irrespective of the precise physical interpretation of $\theta_0^0$,
  for an assumed homogeneous and isotropic spacetime, one would expect it to be coordinate independent because  no wave-particle like duality is involved here. We pointed out that such position dependence actually appears from the coordinate dependence of $g_{\alpha \alpha}$ (for $k \neq 0$). For a model static universe, physical obersables like ``acceleration due to gravity'' $a$ and pressure gradient too would have similar latent coordinate dependence for $k \neq 0$. Thus, we find that though mathematically, constant curvature homogeneous and isotropic spacetime could be of 3 variety,
  physically, only the flat $k=0$ case is admissible.
  
  By considering the empty de-Sitter model, we found that $\theta_0^0$, while coordinate independent, is negative. Further $P_0 = -\infty$ unless
  $\Lambda =0$. Also, we noted that, while in one version of de-Sitter metric, one may see gravitational red-shift, in another version, there would be
  no such gravitational redshift. From all such considerations, we got strong hint that Einstein was right in rejecting $\Lambda$. Such a rejection need not be inconsistent with the observations because there are some analysis which claims that the observations of distant Type Ia supernovae are actually
  in agreement with $\Lambda =0$ picture\cite{30,31}.

  We found that for a supposed static universe, in order that, $P_0$ is non-negative and non-infinite for the closed $k=+1$ case, one should have $S=\infty$. Such a value of $S$ actually tantamount to $k =0$ case because in both the cases curvature $K = k/S^2 = 0$. This again suggests that
  $\Lambda =0$.
 Let us highlight a simple point which might have been overlooked previously. In the Rimennian normal coordinates, where $\partial_c g_{ab} =0$, one expects
  all components of $t_a^b$ to vanish. However as seen from Eq.(6) and as earlier noted by Tolman, in this local Lorentz frame (see Eq.[87.15])\cite{5}
  \b
  t_a^b = {\Lambda\over 8 \pi}  \delta_a^b
  \e
  {\em would not vanish} unless $\Lambda =0$. Thus a finite $\Lambda$ {\em seems to be inconsistent with the principle of equivalence}. 
  
  We found that $\theta_0^0$  contains a dynamic $\ddot S/S$ term; 
  \b
  \theta_0^0 = \theta_0^0 (S, {\ddot S})
  \e
   In one sense, this was
  a welcome feature because otherwise a likely 
  \b
  \theta_0^0 = \theta_0^0 (S)
  \e
  alone would yield the same $\theta_0^0$ for {\em both a  static as well as dynamic} metric and it might appear that motion has no contribution
  to the total energy. However, it turns out that occurrence of this ${\ddot S}/S$ term creates problem for the conservation of $P_0$. In such a case, $P_0$ may vary from $+\infty$ to $-\infty$. It follows then that, in order for $P_0$ to be a physically significant quantity, one should restrict $\rho_* =0$ and $S=\infty$. Incidentally, there are claims that both the distant Type Ia supernovae and Gamma Ray Burst observations may be consistent with a static universe\cite{32,33}. In fact, it might be possible to conceive of static yet continually evolving universe where the patches of matter are infinitely separated from one another\cite{27,28}. Note, when $\rho_* =\Lambda=0$ and $p\ge 0$, we have {\em mean} $\rho =0$.
  However, such a result may be seen  to be at variance with the patch of the universe  {\em observed} now and attendant popular interpretations. But let us pose the question whether  the observed universe is really isotropic and homogeneous at scales say $\sim 100$ Mpc? Well, Sloan Digital Sky Survey shows that there are structures of extent $\sim 500$ Mpc (Sloan Great Wall). Further Wilkinson Microwave Anisotropy Probe (WMAP) has found ``void''
  of the extent of $280$ Mpc\cite{27,28,34}. Thus the observed universe need not be really described by the idealized FRW metric, and one need not be unduly perturbed over the suggestions gleaned from this study. In fact, cosmic background radition may be explained as the redshifted thermal radiation from eternally collapsing objects, i.e., the supposed black hole candidates\cite{27,28}. However, if one would insist that the observed patch of the universe, in any case, must be described by FRW metric,
  then one might adopt two views:
  (i) Einstein's cannonical  pseudo tensor does not provide a physically meaningfully answer to energy momentum conservation. Such a statement would however overlook the fact that when evaluated in quasi-Cartesian coordinates, Einstein EMC does indeed find energy momentum flux for cylindrical gravitational waves\cite{11}. Further, for an isolated static object in AFST, the expression for $P_0$ derived
  from Einstein EMC indeed matches with the same derived by Landau Lifshitz without using any pseudo tensor at all.
  
  (ii) One might also adopt the view that energy conservation is invalid in the cosmological context. Such a view might degenerate into free violational
  of energy conservation principle in astrophysics. For instance, in such a case, while one might see huge eruption of energy or in some cases, one may as well see sudden disappearance of energy. Eruption of energy is indeed seen in astrophysics in the form of supernovae and gamma ray burst and astrophysicists try to understand them using physics which, in the background, honour, energy momentum consevation. And as far as sudden loss or vanishing of energy is considered, no such event has been recorded. In the absence of an energy conservation principle,
motion of cosmic matter too could be unexpected and unpredictable.

Thus, if we would not abandon this principle of energy conservation, there may be examples, how mathematically allowed dynamic motion can be forbidden in GR. For instance, mathematically, one can always formulate equations describing contraction/expansion of clouds without any heat/radiation transfer.
However, an energy conservation principle which involves global gravitational energy dictates that there is no contraction or expansion of self-gravitating objects without heat/radiation transport\cite{24,35}. This principle however may be avoided by assuming that pressure of the fluid $p\equiv 0$, in which case
thermodynamics ceases to work. But a strict $p=0$ equation of state is possible only where $\rho =0$ - -thus such strict $p=0$ solutions would correspond to a fluid mass energy $P_0 =0$ and the mathematical collapse would be devoid of any physical reality.  The symmetry of the FRW metric precludes any heat/radiation flow and this might be forbidding motions even though mathematically the metric suggests motion.

In fact, much earlier to such studies, while studying the simplified problem of the adiabatic collapse/expansion of an {\em uniform} density sphere, Taub found that {\em there would not be any collapse/expansion} if the fluid would have an EOS\cite{36}. Later Mansouri studied this problem with greater clarity
in a paper entitled {\it On the non-existence of time dependent fluid spheres in general relativity obeying an equation of state}\cite{37}. And the basic reason which forbade such motion was principle of energy conservation.  Thus many general possibilities which may be initially suggested by mathematical equations, may not eventually be allowed by rigourous physics.
 Let us recall that, in general, {\em GR does not allow global synchronization of clocks in presence of gravity}, and on the other hand,  clocks are expected to slow down in the presence of gravity.
  Even in the absence of gravity, special relativity does not allow global synchronization of clocks if motion is present. But in the FRW model,
  one expects to have a Newtonian like global time despite the presence of gravity and motion! Does Einstein energy tell that such a global time is
  actually possible only when the model would degenerate into a Newtonian one in the $\rho \to 0$ limit?
  An honest introspection may show that, probably, there were already strong hints that FRW metric, despite its apparent general nature, might 
  be inherently, Newtonian with a mean $\rho \to 0$. This is so because, long back, it was shown  by Milne\cite{38} and McCrea and Milne\cite{39} that the key results of the Friedmann model can be exactly obtained by using pure Newtonian gravity. Nobody has ever explained why it is so! In particular, let us consider the Newtonian
  equation of motion of a test particle   (not acted upon by any pressure gradient or other forces) of mass $m$ lying on the surface of a uniform density
sphere of mass $M$:
\b
F = - {G M m\over S^2}
\e
Since
\b
M = {4\pi \over 3} \rho S^3,
\e
\b
F = m {\ddot S}
\e
one obtains

\b
 {{\ddot S} \over S} = -{4 \pi G\over 3} \rho 
 \e
 Since in Newtonian gravity, $p \sim  \rho/c^2 \ll \rho$, the $3p$ term does not appear in the definition of $M$. But if one is keen, one can introduce $p$ here by an appeal to special relativity\cite{40}. One might also push $\Lambda$ within $\rho$. In any case,  one can see that the FRW evolution Eq.(88) is nothing but the  Newtonian evolution equation! In particular, the $4\pi/3$ factor in Eqs.(88), (135) and (138) may be signalling an inherent $k=0$ geometry. But, we know that GR {\em reduces to exact Newtonian limit} only when $\rho \to 0$. Further when gravity vanishes, the model must be static as found here by demanding conservation of $P_0$. Even if the cosmolgical redshift in the local patch of the universe would  indeed be of dynamical nature, it would be a local phenomenon in such a case because conservation of $P_0$ would mean an infinite static universe where such expanding patches are  separated by infinite voids.
 So let us ponder whether  Einstein energy of the FRW model is eventually explaining the mystery --- why the key equations of the FRW model are obtainable from purely Newtonian physics?
 It is again emphasized that, in any case, observed patch of the universe filled with ``walls'' ``filaments'' and ``voids'' may not be described by the most ideal FRW metric.

 \section{Acknowledgement} The original version of this manuscript focused only on the mere computation of  $\theta_0^0$. But following the suggestion by one of the referees that analysis should be carried out to pursue the physical consequences, the revised version became much more elaborate. Eventually this article  has greatly benefitted from a series of constructive critiques and numerous other suggestions made by this anonymous referee.

\end{document}